\documentclass[12pt]{article}
\usepackage{epsfig}
\usepackage{cite}
\begin{document}
\def\be{\begin{equation}}
\def\bea{\begin{eqnarray}}
\def\ee{\end{equation}}
\def\eea{\end{eqnarray}}
                              \def\barr{\begin{array}}
                              \def\earr{\end{array}}
\def\dis{\displaystyle}
\def\eg{{\em e.g.}}
\def\etc{{\em etc.}}
\def\etal{{\em et al.}}

\def\ie{{\em i.e.}}
\def\viz{{\em viz.}}
\def\lsim{\:\raisebox{-0.5ex}{$\stackrel{\textstyle<}{\sim}$}\:}
\def\gsim{\:\raisebox{-0.5ex}{$\stackrel{\textstyle>}{\sim}$}\:}
                              \def\mev{\: \rm MeV} 
                              \def\gev{\: \rm GeV} 
                              \def\tev{\: \rm TeV} 
                              \def\pb {\: \rm pb}
                              \def\fb {\: \rm fb}
\def\gappeq{\mathrel{\rlap {\raise.5ex\hbox{$>$}}
            {\lower.5ex\hbox{$\sim$}}}}
\def\lappeq{\mathrel{\rlap{\raise.5ex\hbox{$<$}}
            {\lower.5ex\hbox{$\sim$}}}}
\def\ra{\rightarrow}
\def\vevphi{\langle \varphi \rangle}
\def\tilp{\tilde \varphi}
\def\tilh{\tilde h}
\def\tilZ{\tilde z}
\def\tilW{\tilde w}

\vspace*{2ex}

\begin{center}
{\Large\bf Unitarity constraints on the stabilized Randall-Sundrum scenario}
\vskip 15pt

{\sf Debajyoti Choudhury}\footnote{E-mail: debchou@mri.ernet.in}\\

{\em Harish-Chandra Research Institute, \\
Chhatnag Road, Jhusi, Allahabad 211 019, India.}\\

\vspace*{1ex}

{\sf S.~Rai~Choudhury\footnote{E-mail: src@ducos.ernet.in},
Abhinav Gupta\footnote{E-mail: abh@ducos.ernet.in}
and 
Namit Mahajan\footnote{E-mail: nm@ducos.ernet.in}
}

{\em Department of Physics and Astrophysics, 
	University of Delhi, \\ 
	Delhi 110 007, India.}\\

\vskip 15pt
{\Large\bf Abstract}
\end{center}
\vskip 5pt
Recently proposed stabilization mechanism of the Randall-Sundrum
metric gives rise to a scalar radion, which couples 
universally to matter with a weak interaction
($~\simeq~ 1$ TeV) scale. 
Demanding that gauge boson scattering as described by the effective 
low enerrgy theory be unitary upto a given scale 
leads to significant constraints on the mass of such a radion.

\vskip 20pt

\section{Introduction}

The quest to solve the
hierarchy problem that plagues the otherwise successful Standard Model 
(SM) has, over the years, prompted many a plausible extension. Recently, 
it has been proposed that quantum gravity could provide a 
mechanism to stabilize the Higgs mass and thereby
`solve' the problem~\cite{aadd,rs}. Such theories argue that 
if the visible world were 
restricted to a (3 + 1)-dimensional hyper-surface of a larger dimensional
world, then the natural scale for gravity (which propagates in the 
entire bulk) could, conceivably, be as low as ${\cal O}(1-10 \tev)$. 
Amongst these proposals is one by Randall and Sundrum\cite{rs} 
wherein the SM fields live 
on one of the two 3-branes which themselves define the 
ends of the world in the context of a five 
dimensional spacetime. 
The spacetime geometry is nonfactorizable and contains an
exponential warp factor relating the induced metrics on each 
of the two branes. 
This warp factor clearly produces 
a difference in the mass scales between the two end of the world 3-branes
and consequently the (low) natural scale for quantum gravity 
appears, to us, to be incredibly large. 

An issue of particular importance in such models concerns the 
stability of the inter-brane distance (modulus) for this is a key to the 
`solution' of the hierarchy problem. An elegant resolution was 
provided by Goldberger and 
Wise (GW)~\cite{gw}\footnote{For other as well as 
	related proposals of stabilization of the radion modulus 
	see refs.~\protect\cite{stable_others}.}
who introduced a bulk scalar field into the model coupled it 
minimally to the bulk gravity. This simple construction provides 
a nontrivial potential for the radion modulus thereby stabilizing 
it. It was subsequently shown~\cite{cjms} that 
this mechanism, with minor modifications, serves
to stabilize multibrane configurations as well.

The mass of the radion field, given by the 
behaviour of the above mentioned potential close to its minimum, can 
be quite low. Consequently, it could be expected to 
play a nontrivial role in low-energy phenomenology. In fact, so could 
the spin-2 gravitons. The implications of such interactions 
have been examined in the literature quite 
extensively~\cite{phenorefs,RS_unitarity,kim}. 
Not unexpectedly, apart from collider phenomenology, the introduction 
of such a radion as well as the Kaluza-Klein tower of gravitons 
alters the cosmological evolution of the world to such an extent 
that even the familiar Hubble expansion parameter's dependence on matter 
density in the universe differs from the conventional 
one~\cite{Berkl,RS_cosmo}. 
However, subsequent studies~\cite{restoreFRW} have shown that 
a stabilization mechanism, such as the one mentioned above, 
can also serve to reconcile the RS scenario to 
known cosmological observations. 

In this paper we shall strive to examine the role of the radion 
in the context of gauge boson or heavy fermion scattering. It is normally 
expected that, well below the quantum gravity scale, it should be 
possible to treat the RS scenario as
 a field theory, 
albeit a nonrenormalizable one. 
The backbone of the theory is given by a renormalizable gauge theory (the SM)
with the RS character manifesting itself in the form of certain 
additional (and potentially nonrenormalizable) terms in the effective theory. 
Well below the RS scale, then, the theory should look almost unitary. 
This is the aspect that we propose to investigate. Although some 
work has been done in this area~\cite{RS_unitarity}, 
the choice of processes therein was not optimal and hence the 
bounds were rather weak.

In its simplest version, the RS scenario is described by a metric
\be
ds^2 = e^{-2k r_c  |y|}\eta_{\mu\nu}dx^\mu dx^\nu-r_c^2 dy^2 \ ,
	\label{metric}
\ee
where $x^\mu$ are the ordinary 4-dimensional coordinates while 
$y \in [-\pi,\pi]$ parameterizes a $S^1/Z_2$ orbifold. The metric clearly 
describes a slice of $AdS_5$ space with a volume radius $r_c$ and 
a curvature radius $k^{-1}$. For the above metric to be a solution 
of Einstein's equations, the bulk must have  a negative
cosmological constant and the two end-of-the-world branes
at $y=0$ and $y=\pi$ must have positive and negative tension
respectively. An observer on the $y=\pi$ brane 
experiences a red-shift $e^{-k r_c \pi }$ for all its mass parameters
with respect to an observer living at $y=0$. 
Thus, if the $y = \pi$ brane is assumed to be the visible one, 
and if $k r_c\pi \sim 35$,  
the large hierarchy in the ratio $M_{\rm weak}/M_P$ 
could  be explained naturally. 

To be treated as a field theory of gravitation, the metric of
eq.(\ref{metric}) needs to be promoted to space time dependent
fields. The substitution $\eta_{\mu \nu} \ra g_{\mu \nu}(x)$
incorporates both the massless 4-dimensional graviton and its
Kaluza-Klein (KK) counterparts, while $r_c \ra T(x)$ describes the
spin-0 modulus field\footnote{The components $g_{5\mu}$ of the full 
	metric, on KK reduction, result in vector fields which, however, 
	do not couple to the SM fields at the lowest order.}
The volume radius $r_c$ is
thus nothing but the expectation value of the modulus field
$T(x)$. Explaining the hierarchy between the Planck scale and the
electroweak scale thus requires $\langle T(x) \rangle \sim 35/\pi k$. 
Goldberger and Wise achieve this naturally by postulating an extra
bulk scalar field coupled minimally to gravity~\cite{gw}.  The
quantization of the modulus is best done in terms of a redefined field
\be \dis 
\varphi \equiv \vevphi e^{- k \pi (T - r_c)} \qquad \vevphi 
	= \sqrt{\frac{24 M^3}{k}} e^{- k \pi r_c} 
\label{radion} 
\ee 
where $M$ is the Plank mass in the 5-dimensional theory.  
Apart from stabilising
the modulus, the GW potential has the additional consequence that the
mass of the radion field $\varphi$ is much smaller than that of the
lowest lying KK-excitation of the graviton. Thus, in such a scenario,
the radion is more likely to play a significant role in weak-scale
phenomenology than the graviton excitations.

It is easy to see that, at the lowest order, the radion field couples
to the SM matter only through the trace of the energy momentum 
tensor~\cite{gw2,Berkl}. For massive fields, then, the 
relevant interaction 
terms in the effective theory Lagrangian are given by
\be
	{\cal L}_{\rm int} = \frac{\varphi}{\vevphi} T_\mu^\mu
\ee
For massive fermions and vector fields, this reduces to
\be
	{\cal L}_{\rm int} = \frac{\varphi}{\vevphi} 
			\left[ m_\psi \bar \psi \psi 
				+ m^2_V V_\mu V^\mu 
			\right]
	\label{lagrangian}
\ee
The radion coupling to ordinary matter is clearly analogous to that of
the SM Higgs, albeit with a different coupling strength. The radion, thus,
could be looked for in observables wherein the Higgs plays an 
important role. These range from direct production (in associated 
Bjorken process or $gg$ fusion) to radiative effects such as in 
the electroweak precision data. 

A light radion could also signal its presence in $t \bar t$ or vector 
boson scattering. It is well known that, within the SM, the Higgs plays 
an essential role in restoring the perturbative unitarity of such scattering 
processes. With the radion playing the role of an additional Higgs-like 
state, it is conceivable that the extra contribution 
due to a radion exchange could destroy the high-energy behaviour
of such an amplitude. 

Unitarity of gauge boson scattering in the SM has been well 
studied in the literature~\cite{LQT,unit_others}. For longitudinally 
polarized  gauge bosons, the $s$-wave amplitudes are given by
\be
\barr{rcl}
\dis {\cal M}^{({\rm SM})} (Z_L Z_L \rightarrow Z_L Z_L)
   & = & \dis 
     \frac{- i s }{v^2} \: g_{ZZ}(\tilh)
	\\[3ex]
{\cal M}^{({\rm SM})} (W^+_L W^-_L \rightarrow W^+_L W^-_L)
   & = & \dis 
     \frac{- i s }{v^2} \: g_{WW}(\tilh)
	\\[3ex]
{\cal M}^{({\rm SM})} (Z_L Z_L \rightarrow W^+_L W^-_L)
   & = & \dis 
{\cal M}^{({\rm SM})} (W^+_L W^-_L \rightarrow Z_L Z_L) 
   \\[1.5ex]
   & = & \dis
     \frac{- i s }{v^2} \: g_{ZW}(\tilh)
   \\[1.5ex]
v & \equiv & \langle H \rangle \approx 246 \gev
\earr
\ee
where we have neglected terms of ${\cal O}(m_W^2/s, m_Z^2/s)$ and 
\be
\barr{rcl}
g_{ZZ}(x) & = & \dis  \frac{x}{16 \pi}
	\Bigg[ 3 + \frac{x}{1 - x} 
	     - 2 x  
		\ln\left(\frac{1 + x }{x}\right)
	\Bigg] \ ,
	\\[2.5ex]
g_{WW}(x) & = & \dis  \frac{x}{16 \pi} \: 
	\Bigg[ 2 + \frac{x}{1 - x }
	- x \ln\left(\frac{1 + x}{x}\right)
	\Bigg] \ ,
	\\[2.5ex]
g_{ZW}(x) & = & \dis  \frac{x}{16 \pi}\:       \frac{1}{1 - x} 
	\\[2.5ex]
\tilh & = & \dis \frac{m_H^2}{s} 
\earr
\ee
Clearly these amplitudes grow with 
$m_H$ to the extent that a SM Higgs heavier than approximately 
$900 \gev$ renders them nonunitary.

The radion contributions to these amplitudes are quite analogous 
to the Higgs contributions and are easily calculated. 
Concentrating again on the $s$-wave amplitudes, we find that these 
are given by
\be
\barr{rcl}
\dis {\cal M}^{(\varphi)} (Z_L Z_L \rightarrow Z_L Z_L)
   & = & \dis 
     \frac{- i s }{\vevphi^2} \: f_{ZZ}(\tilp)
	\\[3ex]
{\cal M}^{(\varphi)} (W^+_L W^-_L \rightarrow W^+_L W^-_L)
   & = & \dis 
     \frac{- i s }{\vevphi^2} \: f_{WW}(\tilp)
	\\[3ex]
{\cal M}^{(\varphi)} (Z_L Z_L \rightarrow W^+_L W^-_L)
   & = & \dis 
{\cal M}^{(\varphi)} (W^+_L W^-_L \rightarrow Z_L Z_L) 
   \\[1.5ex]
   & = & \dis
     \frac{- i s }{\vevphi^2} \: f_{ZW}(\tilp)
\earr
\ee
where
\be
\barr{rcl}
f_{ZZ}(x) & = & \dis  \frac{1}{16 \pi}
	\Bigg[ \frac{1}{1 - x} 
	     + (2x - 1) 
	- 2 x^2 
		\ln\left(\frac{1 + x }{x}\right)
	\Bigg] \ ,
	\\[2.5ex]
f_{WW}(x) & = & \dis  \frac{1}{16 \pi} \: 
	\Bigg[ \frac{1 + 3 x}
				 { 2 (1 - x) }
	- x^2 \ln\left(\frac{1 + x}{x}\right)
	\Bigg] \ ,
	\\[2.5ex]
f_{ZW}(x) & = & \dis  \frac{1}{16 \pi}\:       \frac{1}{1 - x} 
	\\[2.5ex]
\tilp & = & \dis \frac{m_\varphi^2}{s} 
\earr
\ee
and again terms of ${\cal O}(m_W^2/s, m_Z^2/s)$
have been ignored. 

\begin{figure}[ht]
\vspace*{-4.5cm}
\centerline{
\epsfxsize=9.5cm\epsfysize=11.0cm
                     \epsfbox{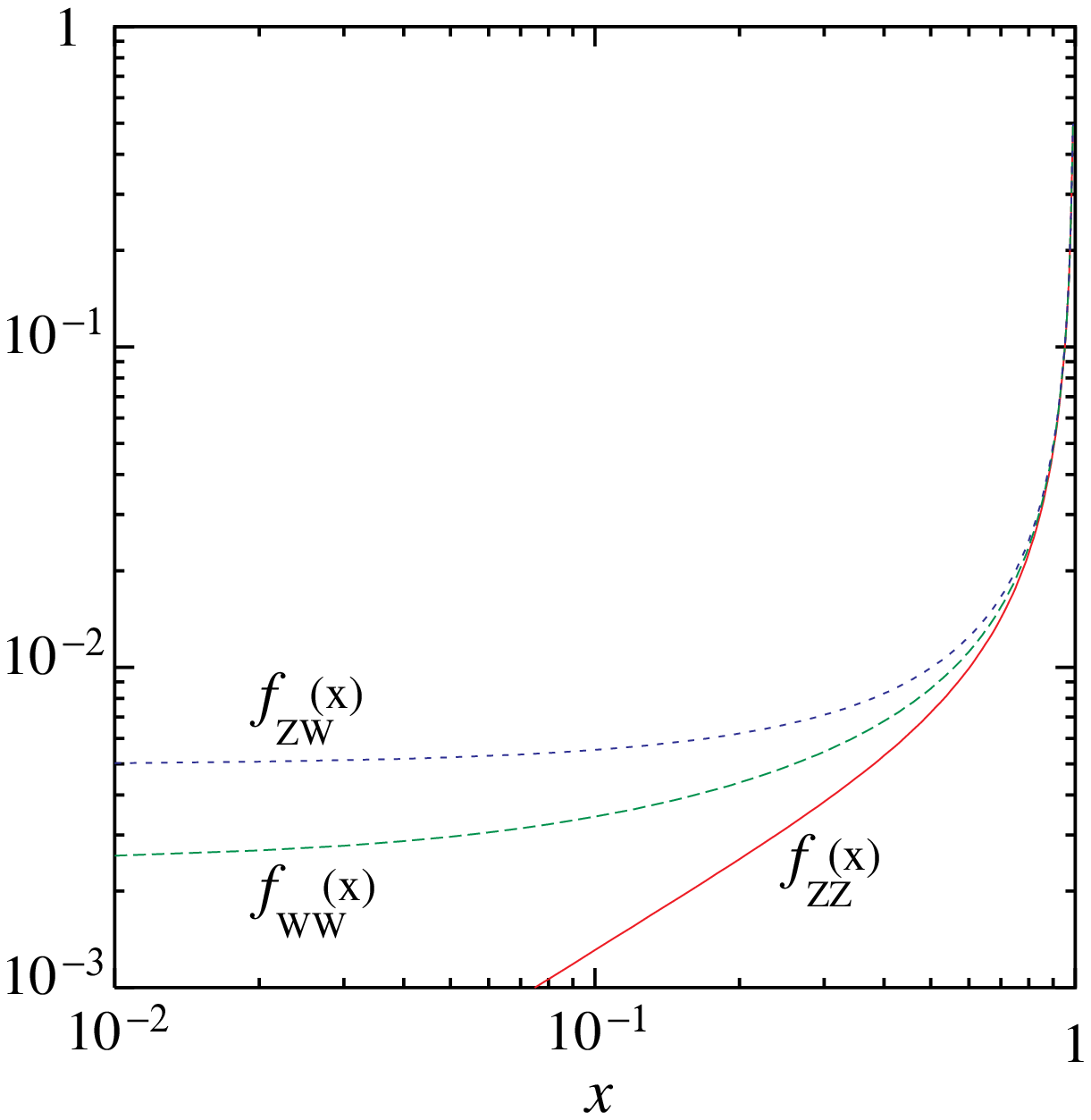}
        \hspace*{-1ex}
\epsfxsize=9.5cm\epsfysize=11.5cm
                     \epsfbox{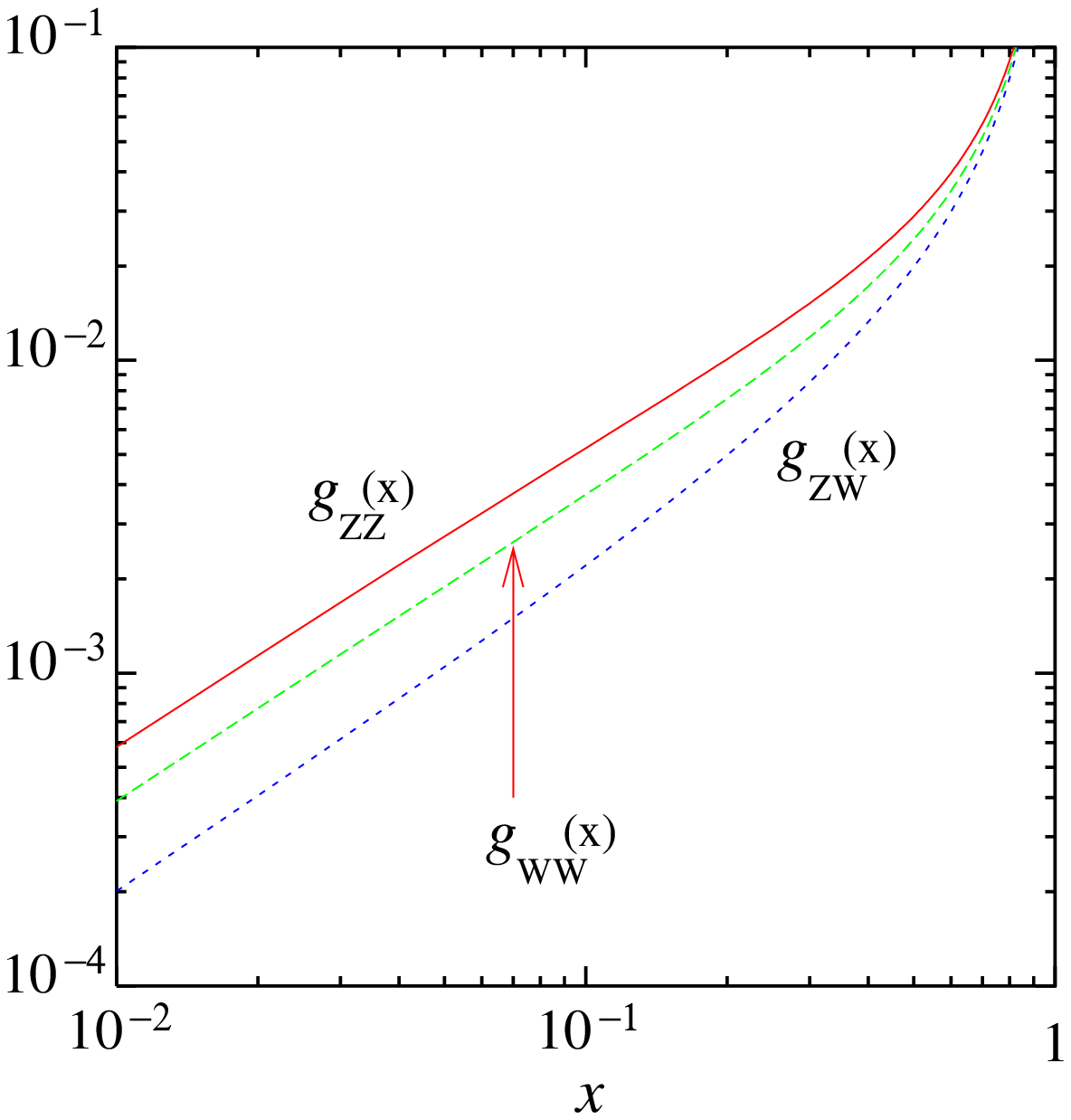}
}

\caption{\em The functions appearing in the expressions for the $J = 0$
        amplitudes. {\bf (left)} radion contribution;
        {\bf (right)} SM contribution.
	}
\label{fig:the_fs}
\end{figure}

We plot the functions $f_i(x)$  and $g_i(x)$ in Fig.~\ref{fig:the_fs}.
The sharp rise as $x \ra 1$ is 
clearly symptomatic of the $s$-channel resonance. 
Beyond $x = 1$, the curves would fall off with the asymptotic 
behaviour being $\sim 1 / x$. However, the region $x > 1$ is 
of no interest to us. A few points need to be noted here
\begin{itemize}
  \item The SM amplitudes and the radion contributions have the 
	{\em same sign} and hence there is no scope of destructive 
	interference.
  \item For large $x$ (i.e. $x \lappeq 1$), the functions $g_i(x)$
	typically dominate $f_i(x)$. To this information, 
	one has to couple the fact that 
	the pre-factor $\vevphi^{-2}$ is  expected to be smaller
	than $v^{-2}$. Thus, for similar Higgs and radion masses,
	it is obvious that the SM amplitudes 
	are bigger, for this range of $\sqrt{s}$, 
	than the radion contributions. 
	In other words, unitarity bounds, if any, would be primarily 
	driven by the SM dynamics. 
  \item For small $x$, on the other hand, $f_i(x)$ easily dominate 
	$g_i(x)$ and the situation regarding possible unitarity 
	bounds gets reversed. This is but a reflection of the fact
	that, for a light Higgs, partial wave unitarity is respected
	by the SM, while it might be suspect in the theory with radions.
  \item For small x, 
	$f_{WW} (x)$ and $f_{ZW} (x)$ reach constant values of $1/(32\pi)$ 
	and 
	$1/(16 \pi)$ respectively, while $f_{ZZ}(x)$ falls off as 
	$3 x / (16 \pi)$. Hence, for a given radion mass $m_\varphi$, 
	the last three amplitudes grows with the center of mass energy
	$\sqrt{s}$ and thus stand to violate partial wave unitarity.
	The amplitude $Z_L Z_L \rightarrow Z_L Z_L$ though 
	goes over to a constant value.
\end{itemize}

Looking at the arguments listed above, it becomes clear that the 
{\em most conservative} bounds on the radion parameter space would emanate 
for a light Higgs. In our analysis, therefore, we shall consider 
the smallest mass allowed to the SM Higgs, namely $114 \gev$. In fact, 
there are even some preliminary signals of a Higgs with a 
very similar mass~\cite{higgsmass}. Although it is only the 
${\cal M}^{(\varphi)} (W^+_L W^-_L \rightarrow W^+_L W^-_L)$
and 
${\cal M}^{(\varphi)} (Z_L Z_L \rightarrow W^+_L W^-_L)$ that are 
expected to give the strongest bounds, one must remember that 
these channels are coupled.\\
For completeness, we consider not only coupled $W_LW_L$, $Z_LZ_L$
channels but also couple them to $hh$-channel. To this end,
the complete set of $J=0$ partial wave amplitudes is:
\be
\barr{rcl}

\dis {\cal M}(hh \rightarrow hh) &=& \dis \frac{-\iota}{16\pi\vevphi^2s}
\Bigg[ \left(3m_{\varphi}^2 + 16m_h^2 +  
\frac{(2m_h^2 + m_{\varphi}^2)^2}{s - m_{\varphi}^2}\right)~(s - 4m_h^2) \\ 
&-& 2(2m_h^2 + m_{\varphi}^2)^2~\ln\left(\frac{s + m_{\varphi}^2 - 4m_h^2}
{m_{\varphi}^2}\right) \Bigg] \\ [3.0ex]

{\cal M}(Z_LZ_L \rightarrow Z_LZ_L) &=& \dis \frac{-\iota}{16\pi\vevphi^2s}
\Bigg[ \left(3m_{\varphi}^2 - 8M_Z^2 + \frac{(m_{\varphi}^2 - 2M_Z^2)^2}
{s - m_{\varphi}^2}\right)~(s - 4M_Z^2) \\
&-& 2(m_{\varphi}^2 - 2M_Z^2)^2~\ln\left(\frac{s + m_{\varphi}^2 - 4M_Z^2}
{m_{\varphi}^2}\right) \Bigg] \\ [3.0ex]

{\cal M}(W_LW_L \rightarrow W_LW_L) &=& \dis \frac{-\iota}{16\pi\vevphi^2s}
\Bigg[ \left(s + 2m_{\varphi}^2 - 8M_W^2 + \frac{(m_{\varphi}^2 - 2M_W^2)^2}
{s - m_{\varphi}^2}\right)~(s - 4M_W^2) \\
&-& \frac{1}{2}(s - 4M_W^2)^2 -  
(m_{\varphi}^2 - 2M_W^2)^2~\ln\left(\frac{s + m_{\varphi}^2 - 4M_W^2}
{m_{\varphi}^2}\right) \Bigg] \\ [3.0ex]

{\cal M}(Z_LZ_L \rightarrow W_LW_L) &=& \dis
{\cal M}(W_LW_L \rightarrow Z_LZ_L)\\ &=& \dis
\frac{-\iota}{16\pi\vevphi^2s}
\Bigg[\frac{(s - 2M_Z^2)(s - 2M_W^2)}{s - m_{\varphi}^2} 
[(s - 4M_Z^2)(s - 4M_W^2)]^{\frac{1}{2}}\Bigg] \\ [3.0ex]

{\cal M}(hh \rightarrow V_LV_L) &=& \dis {\cal M}(V_LV_L \rightarrow hh)\\
&=& \dis \frac{-\iota}{16\pi\vevphi^2s}\Bigg[
\frac{(s - 2M_V^2)(s + 2m_h^2)}{s - m_{\varphi}^2} 
[(s - 4M_V^2)(s - 4m_h^2)]^{\frac{1}{2}}\Bigg]
\earr
\ee
with $V ~=~ Z ~or~ W$.\\

\begin{figure}[ht]
\vspace*{-3.5cm}
\centerline{
\epsfxsize=11.5cm\epsfysize=13.0cm
                     \rotatebox{270}{\epsfbox{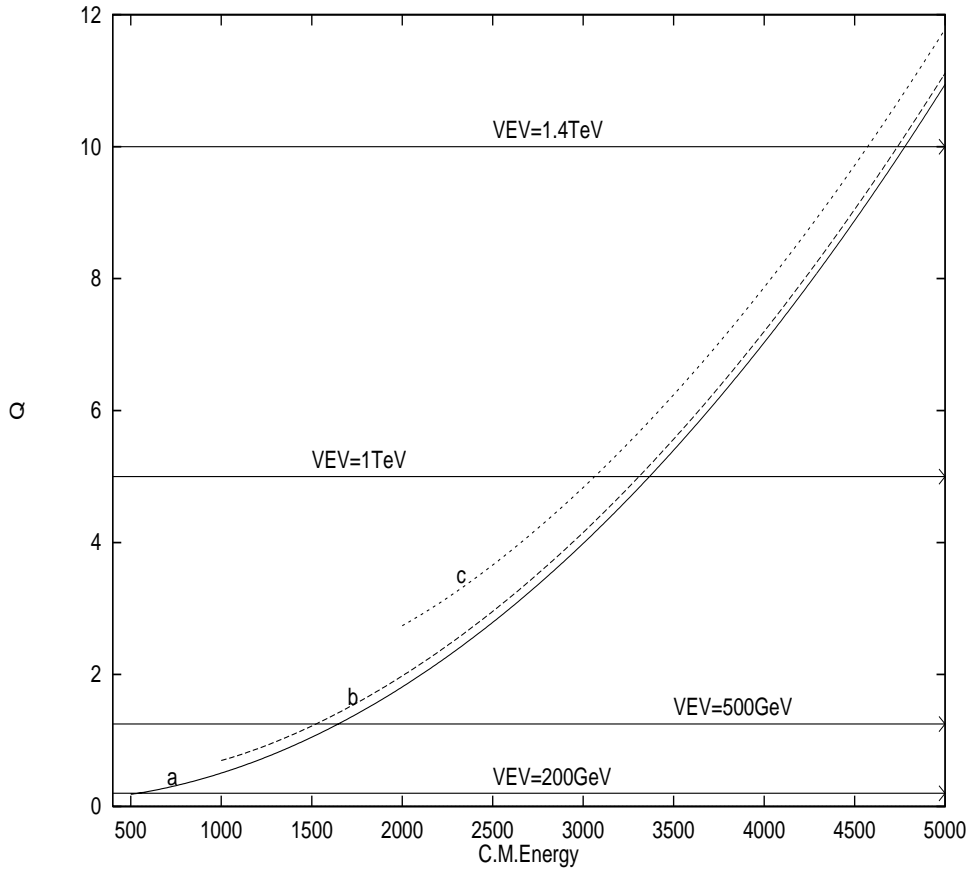}}
}

\caption{\em $Q~=~\vevphi^2|\lambda_{max}| \times 10^{-5}~ (GeV)^2$ vs 
             $\sqrt{s}~ (GeV)$
             curves for different choices of $m_\varphi$. Curves a, b and c
            correspond
            to $m_\varphi =$ 250, 500 and 1000 GeV respectively.  
	}
\label{fig:comp}
\end{figure}

Hence, we need to consider the eigenvalues of the matrix 
\[
	\pmatrix{{\cal M}_{WW} & {\cal M}_{WZ} & {\cal M}_{Wh} \cr
		 {\cal M}_{ZW} & {\cal M}_{ZZ} & {\cal M}_{Zh} \cr
                 {\cal M}_{hW} & {\cal M}_{hZ} & {\cal M}_{hh} \cr
		}
\]
where the amplitudes ${\cal M}_{ij}$ follow an obvious notation. \\
These amplitudes are all real and hence the unitarity constraint on these
amounts to demanding that the magnitude of the highest eigenvalue,
$\lambda_{max}$ satisfy:
\be
|\lambda_{max}| \le \frac{1}{2} \ .
	\label{unitar_condn}
\ee
Although the matrix elements obviously contain the full amplitudes 
(SM as well as radion contributions), it turns out that neglecting the 
SM contribution is a good approximation, particularly for 
small Higgs masses.\\ 
Before we actually embark on analysing the unitarity bounds, 
let us, briefly, consider these eigenvalues. These are obviously 
functions of the three independent 
quantities $\sqrt{s}$, $m_\varphi$ and $\vevphi$.
However, in the approximation that we are working in, the eigenvalues are 
dependent on $\vevphi$ only by a direct multiplicative factor $\vevphi$.
Our results thus can be conveniently plotted as $\vevphi^2|\lambda_{max}|$
as a function of $\sqrt{s}$ for various choices of $m_\varphi$ and 
these are shown in Fig.~\ref{fig:comp}.
 The unitarity bounds for various choices of
$\vevphi^2$ are then horizontal lines as shown therein.\\ \\ 
\vskip 3.0cm
A very relevant feature to benoted for the present problem is that 
the amplitudes and hence the eigenvalues, increase asymptotically with
$\sqrt{s}$. This is unlike the SM case where because of cancellations between
various contributing graphs, the amplitudes approached constant values at
large $\sqrt{s}$ but proportional to $m_h^2$. This last feature allowed one 
to put bounds on $m_h^2$ by putting perterbative unitarity restrictions
on the amplitudes. We cannot directly follow the same procedure here but
use a procedure followed in \cite{hasenfratz}. For an effective theory to 
be reasonable, it should be valid at least till energies comparable and 
somewhat above the particle masses in the theory. As a rule then,
we can demand that perterbative unitarity be valid till energies $\sqrt{s}$
equal to $2m_\varphi$. With this, we see that if the radion coupling is
weak, e.g. $\vevphi^2 \sim 2~(TeV)^2$, no violation is seen even for very
heavy radion mass $\sim 1$ TeV and thus no meaningful limit is obtained.
At the other end, for strongly coupled radion as has been considered in
\cite{abc}, typically $\vevphi = 200$ GeV, there will be a limit on 
$m_\varphi$, about 300 GeV, above which the conditions discussed above will
be violated. at intermediate $\vevphi$, typically $\vevphi = 500$ GeV,
the corresponding limit on $m_\varphi$ becomes higher.\\
Admittedly, the conclusions reached here do not have a very definitive
character as compared to the one 
reached in \cite{LQT} because unlike the VEV of the higgs field, the 
corresponding radion VEV is unknown. The energies upto which the unitarity
restrictions are not violated represent an effective upper limit of
energies for which the underlying theory/interactions can be considered
as an ``effective theory''.    \\
We make a brief digression here to consider other possible 
interactions wherein radion exchange may play a significant role. 
Clearly any such process should involve only heavy particles. 
Apart from the gauge bosons (and the Higgs), the only other heavy particle
within the SM is the top quark.
As pointed out earlier, a process involving $t\bar{t}$ could also 
serve to be a signal for the presence of the radion. 
One such process is $Z_L Z_L \rightarrow t \bar{t}$. 
The radion contribution to 
$J = 0$ amplitude for the $+ +$
($= - -$) helicities of the $t\bar{t}$ final state is given by
\be
\barr{rcl}
\dis {\cal M} (Z_L Z_L \rightarrow t \bar{t})
   & = & \dis 
     \frac{i m_t \sqrt{s} }{16\pi \vevphi^2} \:
	\frac{1}{1 - \tilp} 
\earr
\ee
while the cross helicity amplitudes vanish. 
The high energy behaviour for
 $Z_L Z_L \rightarrow t \bar{t}$ is better behaved as compared to 
the same for gauge-boson scattering, thus leading
to much weaker constraints. A similar statement holds for 
$W_L^+ W_L^- \rightarrow t \bar{t}$ as well.

Equation (\ref{unitar_condn})  imposes an inequality in a
space spanned by $\sqrt{s}$, $m_\varphi$ and $\vevphi$. 
Note that while the authors of Ref.~\cite{hasenfratz} 
prefer $x_c = 1/2$, we have chosen to be more general.\\ \\ 
\vskip 2.5cm
\begin{figure}[ht]
\vskip 1.5cm
\vspace*{-4cm}
\centerline{
\epsfxsize=11.5cm\epsfysize=8.5cm
                     \epsfbox{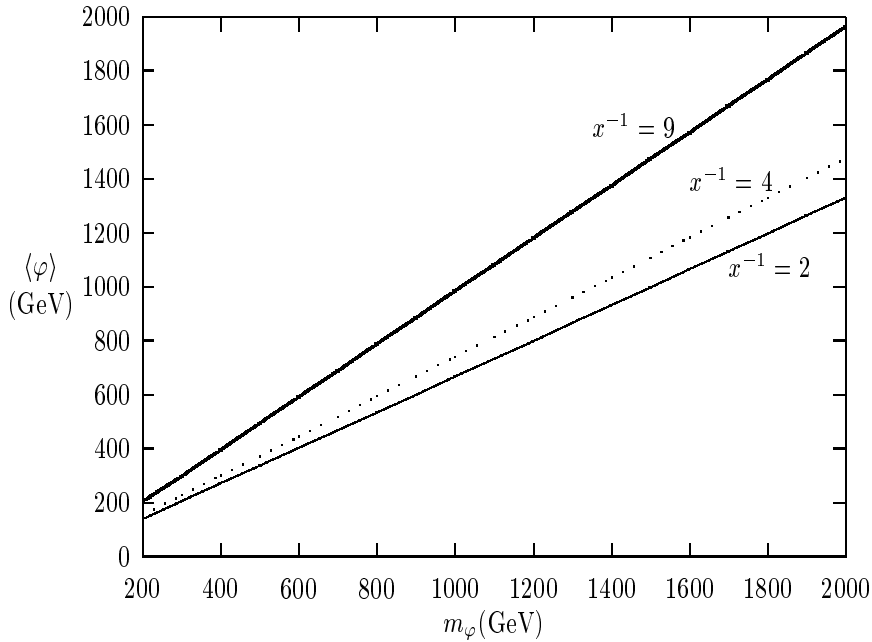}
}

\caption{\em Constraints on the parameter space obtained by demanding 
	that the $J = 0 $ amplitudes for longitudinal gauge-boson scattering
	respect unitarity bounds upto an energy scale given by 
	$s = x^{-1} m_\varphi^2$. The region {\em above} the curves are ruled 
	out.}
	\label{fig:constraint}
\end{figure}

In conclusion, the present investigation,
 unlike the parallel one for Higgs mass, thus yields
no bound on the radion mass but only constrains 
the `allowed' region in the  $m_\varphi$--$\vevphi$ plane. 
However, what makes this result a little more 
interesting is the fact that this allowed region is somewhat complimentary
to the ones obtained by Kim et.al~\cite{kim}, in their investigation relating
to neutral current data. Since only the intersection of such 
allowed domains is truly permissible, the present study could prove
rather useful in defining future search strategies for the
radion.

\vspace*{2ex}
{\bf {\em Acknowledgements:}} DC would like to thank the Department of 
Science and Technology, India for financial assistance under 
the Swarnajaynti Fellowship grant. AG would like to thank C.S.I.R.,
India and NM thanks University Grants Commission, India for fellowships.

\vspace*{2ex}

{\em Note added:} As this manuscript was being finalised, 
a very similar work~\cite{han} was published. Although the two 
papers share a few common points, the primary focus are somewhat 
different.

\end{document}